# A QUANTUM-THEORETICAL APPROACH TO THE PHENOMENON OF DIRECTED MUTATIONS IN BACTERIA
(hypothesis)

**Vasily V. Ogryzko**

Laboratory of Molecular Growth Regulation,
NICHD, NIH, Bethesda, Maryland, 20892, USA

**Summary**

Darwinian paradigm of biological evolution is based on the independence of genetic variations from selection which occurs afterwards. However, according to the phenomenon of directed mutations, some genetic variations occur mostly when the conditions favorable for their growth are created. I propose that the explanation of this phenomenon should not rely on any special "mechanism" for the appearance of directed mutations, but rather should be based on the principles of quantum theory. I consider a physical model of adaptation whereby a polarized photon, passing through a polarizer, changes its polarization according to the angle of the polarizer. This adaptation occurs by selection of the "fitted" polarized state which exists as a component of superposition in the initial state of photon. However, since the same state of the incoming photon should be decomposed differently depending on the angle of the polarizer, in this case the set of variations subjected to selection depends upon the selective conditions themselves. This reveals the crucial difference between this model of adaptation and canonical Darwinian selection.
Based on this analogy, the capacity of cell to grow in particular conditions is considered an observable of the cell; the plating experiments are interpreted as measurement of this observable. The only nontrivial suggestion of the paper states that the cell, analogously to the polarized photon, may be in a state of *superposition* of eigenfunctions of the operator which represents this observable, and with some probability can appear as a mutant upon the measurement. Alternative growth conditions correspond to the decomposition of the same state vector into different superposition, consistent with measurement of different observable and appearance of different mutants. Thus, consistent with the suggested analogy, directed mutations are explained as a result of random choice from the set of outcomes determined by environment.

## 1. Introduction.

A number of recent papers have been devoted to the question: "To what extent are the genomic changes of bacteria random and non-directed?". Their results once again bring the biologists' attention to the problem of Lamarckism.

For example, Cairns et all (1988) reproduced the results of the fluctuation test first observed 30 years ago by Rian (1952). Rian's experiments could be interpreted as a demonstration of induced appearance of E.coli Lac-revertants able to metabolize lactose in response to its addition to growth medium. In the work of Cairns et al, control experiments were made which eliminated some trivial explanations of this observation (Fig.1). The evidences supporting this phenomenon were observed in other experimental systems as well (Shapiro 1984, Hall 1988). Apart from point mutations they include transpositional changes and deletions. The conclusions of these experiments undoubtedly contradict current ideas about evolutionary mechanisms, "the Central Dogma of Molecular Biology", and provoked an animated discussion in *Nature* (Charlesworth,D. et al 1988, Symonds,N. (1989), Lenski,R. et al, *Nature* (1989)) and dozens of follow-up publications. These results indicate that a bacterium "knows" in some unexplained way how it should change its genome for optimal propagation.

This phenomenon was called the "directed mutations phenomenon". Since its rediscovery, various mechanisms have been proposed from involvement of reverse transcription (Cairns 1988) to the enhanced mutability of transcriptionally active gene (Devis 1989) to the existence of a short living hypermutable states of the cell (Hall ).

Despite the fact that a number of authors doubt the reality of this phenomenon (Charlesworth,D. et al 1988, Symonds,N. 1989, Lenski,R. et al,1989), recent papers by Foster et all (1994) and Rosenberg et all (1994) should convince even the skeptics that directed mutations do occur. Although the mechanism underlying this phenomena is not evident from these studies, they nevertheless point on the difference in the sources of genetic variability between stationary and exponentially growing cells. Finally, the ability to directly alter genome has been found in yeast as well (Hall 1992b; Steele, Jinks-Robertson 1992), indicating a broader applicability of the phenomenon.

For more detailed discussion of the experimental systems and proposed mechanisms the readers are referred to several reviews (Foster 1992, Lenski 1993, Hall 1994). The purpose of this paper is to suggest another approach to the phenomenon of directed mutations. Instead of proposing ad hoc some specific molecular mechanisms for appearance of directed mutations, this approach is rather based on general considerations concerning the phenomena of adaptation and life.

Consider a simple experiment described by Cairns. Cells are plated on lactose containing agar and appearance of mutant colonies is observed. Let's change our perspective from "cell-centric" to "medium-centric". In terms of physical laws the system "cell + medium" may be regarded as a closed system. As for any closed system this one "*strives*" to reduce its free energy. This happens if the cells propagate and utilize lactose. Thus, the appearance of mutants is *energetically beneficial* for the whole "cell + medium" system.

Is it possible to consider this "purposefulness", that is contained in the very basis of physics, the driving force behind directed mutagenesis? Is the ability to mutate in a directed manner, in fact, a natural (though overlooked) property of cell propagation, considered above as dynamic of the whole "cell-medium" system?

Almost all that molecular biology tells us about the intracellular processes contradicts the existence of special feedback mechanisms which would decide whether wrong protein is useful and transfer this information "from protein to gene". But the phenomenon we are interested in is being observed phenotypically, i.e. on the living cell as a whole. Besides, we must be aware of the fact that in spite of the vast volume of knowledge gained by molecular biology, *almost nothing is known about the processes occurring in a single cell*. For example, the majority of facts of the intracellular dynamics in vivo have been obtained while studying 10 million cells at the minimum. It is not always correct to project these results to a single cell by averaging. I believe that a switch to examining "elementary living object" will lead to the important changes in our ideas about intracellular processes.

In this paper, however, I will propose a theoretical approach to the directed mutations phenomenon that will not address the subject of intracellular dynamics. For that I make use of Niels Bohr's supposition that the description of the living object will require complementary and in a sense incompatible approaches (Bohr 1958). We shall point to the limits in which the future of a single cell can be predicted. I shall propose the language which takes these limits into account and which is better for the description of Cairns' experiments. It will be the language of quantum theory. The existence of the directed mutations phenomenon is natural in the framework of such a description.

## 2. Can the fate of a single bacterial cell be predicted ?

In this section I will discuss the following thesis: The genomic sequence of a cell and its capability to propagate in certain conditions are two different properties of the cell, which cannot be determined for a *single cell* simultaneously.

1. It is common knowledge that when cells are plated on selective medium, some of them will form a colony. The colony-forming cells usually are referred to as mutants, due to changes in the genomic structure. According to the phenomenon of directed mutations, not all of the mutants preexist in the initial population before plating. Therefore we can consider a cell as a mutant with complete confidence only *post-factum*. The mutability implies that we can not predict the future of every *single cell* upon plating. Only statistical predictions concerning a large amount of cells may be precise enough.

2. Any attempt to determine more precisely the genomic sequence of a cell will influence the cell's functioning, affecting its capacity for propagating. For example, modern gene technology allows us to extract DNA from a single cell (Erlich,H., Arnheim,A. 1992), then amplify the gene of interest by PCR, determine the nucleotide sequence and conclude whether the cell was mutated or not. The possibility for DNA-polymerase to make error during the first cycle of PCR makes this method less than 100% accurate; one can imagine more precise way to do it (using scanning tunneling microscopy of a single molecule of DNA). Both these approaches, however, entail disruption of the cell and complete loss of the object of investigation.

One can not exclude in principle a possibility of defining genome structure without killing the cell. We, however, will loose in precision. More important, any method of this kind allows us to see only *already existing mutations* and cannot therefore help us with mutations that have emerged as a result of interaction with the medium (e.g., directed mutations). Appearance of directed mutations is a rare event; its frequency is very small, being comparable with that of regular spontaneous mutations ($10^{-6}$-$10^{-7}$). Hence, the appearance of directed mutations is based on chance event(s). It is very likely, therefore, that any attempt to predict the fate of a cell for this case entails meddling in the delicate process of cell-medium interaction and will ultimately affect the results of this interaction.

3. On the other hand, instead of determining the genomic sequence of the cell, we may choose to just let the cell propagate itself and allow the events to develop in their own way. This will result in the irreversible loss of the initial object, (i.e. the single bacterial cell), and will make it impossible for us to know its genomic sequence. Determination of the genomic sequence of the daughter cells can not help us, due to the possibility that the colony has emerged as a result of mutation.

Thus, on the other hand, the *capability to propagate under given conditions* is a property that also can be measured only at the cost of the irreversible change of the experimental situation.

4. Similar restrictions on possibilities of observation form the basis of the quantum theory, which has lead to new ideas about causality different from those concluded from the macroscopic observations. These ideas form the foundation of the complementary principle formulated by Niels Bohr. He expanded them to the biological objects (Bohr 1958) as well. Holevo (1982), analyzing the measurement problem in quantum theory, also pointed out that the "nonclacissity" of quantum-mechanical observables is not necessarily based on the microscopic character of the objects of investigation. The impossibility of simultaneous measurements of different observables of the object is more important. This situation, according to Holevo, may emerge when studying the functioning of single living cells: measurement of one observable may result in changes in the state of the system and thereby affect the outcome of other measurements.

My approach may be regarded as an application of these ideas to the explanation of the directed mutations phenomenon.

## 3. Measurement, operators and superposition.

The first step in the proposed approach is the following suggestion. Describing the experiment with bacteria plating, we will consider it as a measurement procedure and use quantum-mechanical formalism for its description. The implications of this are following. We will consider the cell as a natural object and *the capability to propagate in specific conditions as its observable* which is being measured in the plating experiment. In this description, the Petri dish with selective medium assumes the role of the measurement device. It is a macro-object in a metastable state which is capable, after interaction with the cell, to change irreversibly into one or another final state (either the presence or absence of colony).

The reasons to consider the bacteria-plating as a procedure corresponding to a quantum-mechanical measurement are following:

1. The irreversible character of a bacterium and nutrient medium interaction and unpredictable, in principle, (see 2-1) outcome of the process.

2. This procedure is a measurement by the way it naturally is - we do it to determine some characteristic of the explored object.

3. There is an analogy between the quantum theory's descriptions of the processes of measurement and self-propagation noted by Wigner. I can briefly express its essence as follows. In both cases a "small" system (either measured object or self-reproducing one) interacts with a "large" system (either device or external medium). As a result some changes are induced in the "large system". These changes are, virtually, an amplification of the information which corresponds to the "small" system. According to Wigner, literally, "each [measurable state] is a reproducing system" (Wigner 1963). This allows me, in turn, to consider real self-reproduction as a measurement procedure.

The main features of the description proposed are the following (see, e.g. Dirac (1947)):

Each observable in the quantum theory is represented by a hermitian operator, defined in the space of states of the described system. Herein, for example, "the capability to propagate in lactose" will be represented by an operator that we will call "*Lac*".

The capability to propagate is a property of the whole cell and can not be reduced to any smaller fraction. So it is important to note that any operator of this type acts on the state vector of *the whole cell*. (The state vector describes with maximum precision any physical object, even the cell, that may be ultimately regarded as a system of a large number of nuclei & electrons.).

Operator "*Lac*" singles out the so called *eigenfunctions* from all possible state vectors of the system. The eigenfunctions are not changed upon action of our hypothetical operator; they can only be multiplied by a constant. This property of eigenfunctions reflects the essential role of operator formalism in the proposed approach; it allows us to describe, in this way, a choice of states consistent with given experimental conditions and corresponding to different outcomes of measurement.

There are three eigenfunctions in our case: $|\psi_1\rangle$ corresponds to cell death $|\psi_2\rangle$- to the stationary state, $|\psi_3\rangle$ - to the self-reproduction. (I have simplified the situation. There may be many possible variants leading to growth, each with its own eigenfunctions ($|\psi_4\rangle$, $|\psi_5\rangle$etc). But it is sufficient for the explanation to consider $|\psi_3\rangle$ to be the only one able to grow in the given conditions).

The next step of our approach is the key step and relies on the following nonclassical feature of quantum theory: an object may be in a state described by the *superposition* of different eigenfunctions of given operator. Thus, I suggest the following *hypothesis of superpositon*: the state vector of a cell may be represented as:

$$|\Psi\rangle = c_1|\psi_1\rangle + c_2|\psi_2\rangle + c_3|\psi_3\rangle \qquad (1)$$

with ci as the coefficients related to the probability of the corresponding results of the measurement.

Operationally, this suggestion implies that each cell with non-mutant genome has a certain probability of becoming a colony after being plated (i.e. of manifesting itself as a mutant cell while being "measured"). The assumption of superposition is the only nontrivial suggestion in this work and lays in the core of the hypothesis proposed in the article. Conceptually, it is based on the limitations on simultaneous determination of the genomic sequence of a cell and of its capability to propagate in given conditions (see section 2). As will be seen in the next section, the phenomenon of directed mutations naturally follows from this assumption within the framework of proposed formalism.

It must be pointed out that the proposed idea of *superposition does not contradict common sense* inasmuch as the states forming the superposition are in principle *indiscernible* in this case. They become discernable only upon measurement. This is the point of principal distinction from the famous "Schroedinger cat" (Shroedinger 1928) (see 6.3).

## 4. Quantum-mechanical model of adaptation.

Prior to describing how the approach works, I will consider a quantum-mechanical model of adaptation, which demonstrates important distinction between concepts of "pure" and "mixed" state used in quantum theory. "Pure state" refers to a state, described by state vector, (e.g., $|\Psi\rangle$). When describing a measurement, $|\Psi\rangle$ can be always represented as a superposition of eigenfunctions of relevant operator, reflecting uncertainty in outcomes of the measurement. In addition to superposition, quantum theory has another way to introduce uncertainty in the state of an experimental system. It is the concept of mixed state, which takes into account additional uncertainty in our knowledge of what is pure state of the system. For proper description of mixed state, density matrix approach is necessary ( Bohm 1986 ).

According to the proposed formalism, the concept of "*pure state*" describes a cell with definite genetic sequence; and the directed mutations emerge by measurement of this state. The following quantum-mechanical model of "adaptation" illustrates how this works.

It is well known that a polarized photon can, with a certain probability, pass through a polarizer turned to some angle with respect to the photon polarization. If it passes, its polarization becomes correspondent to the polarizer's one; we will call it "*adaptation of polarization*". One may say that the way this adaptation occurs is due to selection of relevant polarization state that in some sense existed in the state of incoming photon; therefore this is process similar to the darwinian picture of elementary evolutionary process. However, there is the following essential difference. It is incorrect to say that the photon has passed through the polarizer because of a *random* change of polarization to the relevant angle *prior to interaction*. This situation may be called "selection from mixed state"; this would be the model for the "neo-Darwinian" picture of elementary evolutionary process, with the mutants spontaneously generated prior to selection. However, according to the quantum theory, the eigenfunction selected does not exist separately before the measurement, it is "part" of the pure state. What is important in this distinction between the "neo-Darwinian" and the quantum description is that another angle of the polarizer would correspond to different decomposition of the same state vector $|\Psi>$, and lead to the different outcomes of measurement. It is the specific experimental situation (which implies interaction with the device) that makes us to decompose the $|\Psi>$ into superposition of relevant orthogonal basis states (e.g. $|\psi_1>$ and $|\psi_2>$) and to evaluate the fraction of the state that will pass through the polarizer (e.g., $|\psi_1>$).

This picture reflects the well-known active role of the device in quantum-mechanical measurement. One may say that in quantum *theory chance always has an "adaptive" character*: the reduction of a state vector occurs with respect to certain basis (the basis is determined, for example, by measuring device); therefore, the reduction is a random choice, but what is the set of outcomes is determined by the conditions of measurement.

The use of the quantum theory formalism in our approach is based on idea that the environment plays a similar role in many instances of biological adaptation, in particular, in the case of directed mutations. According to current understanding (Zurek 1993), the so called "preferred basis" for the description of macroscopic system consists of the states of the system which "survive" the interaction with the environment, which is considered as constantly "observing, measuring" the system. Upon the change in environment, the preferred basis can also change. In this case the states of the old preferred basis will have to be presented as linear combinations (superpositions) of the new basis. Then, the "adaptation" of the system to the new environment will occur as a reduction of the old state vector to one of the new ones. In other words, the state vector reduction describes how a previously stable state of the system (corresponding to one of the preferred states), becomes unstable (or metastable) upon a change in the environment and has to be resolved to a new stable state.

**5. Adaptive mutations as the result of random choice from set of outcomes determined by environment**.

According to the phenomenon of adaptive mutations, they do not accumulate just in the starving cells, indicating that the application of selective medium actively causes their emergence. This was repeatedly demonstrated in control experiments, when cells were incubated without substrate (i.e. lactose) for several days, after that the substrate was added to the plate (Fig.1B). The time course of appearance of the colonies indicated that the starting point of accumulation of mutants coincided with the substrate addition, with practically no mutants preexisting before. This rules out starvation as the source of the mutations since, in this case, the preexisting mutants would be expected to accumulate before the addition of the substrate. (For more controls and detailed discussion see original papers by Cairns (1988) and reviews by Foster (1992) and Hall (1994)). Thus, the selective medium does not just play the role of a passive detector of the mutants, it is also causing their very emergence, which, in fact is the phenomenon of directed mutations.

It is natural to ask, how selective medium can increase the probability of mutations to occur. However, as the discussion in the previous section suggests, this way to formulate the question is misleading. According to the proposed approach, it is not the change of probabilities of events, but *rather the change of the very set of events that makes the phenomenon adaptive.* Say that we measure a photon polarization by placing a polarizer between the source of photon and the detection screen (Fig.2A). As one can see, the change in the conditions of measurement (change of angle of polarizer) does not affect just probability of detection to happen, but rather the set of outcomes (the photon which passes the polarizer will have different polarization). In fact, if the incoming photon had circular polarization, the probability to pass through the polarizer will not change at all. Nevertheless, this process will still be an adaptation, according to the analogy suggested in the previous section.

Implicit in the formalism proposed in this work (and consistent with the idea of nonseparability in the cell, alluded to in the section 6.3) is that mutation can not be reduced to just a change on the DNA level and should be considered as an event occurring on the level of the whole cell. Lets now, following this idea, instead of a photon take a cell and also substitute a selective medium for a polarizer (Fig2.B). Then, in order to obtain the phenomenon of adaptive mutations, it is sufficient to suggest that, depending of what are the measurement conditions, the state of the cell should be decomposed into a linear combination of eigenfunctions of corresponding operator:

$$|\Psi> = c_1|\psi_1> + c_2|\psi_2> + c_3|\psi_3> \qquad (1)$$

with the mutant $c_3|\psi_3>$ having small contribution $c_3$, reflecting small probability to be detected. Another selective medium will correspond to different operator and therefore to different decomposition of the same state vector, say:

$$|\Phi\rangle = b_1|\phi_1\rangle + b_2|\phi_2\rangle + b_3|\phi_3\rangle \qquad (2)$$

Indeed, we have essentially the same situation as in the previous section: depending on what the selective medium is (angle of polarizer), we will get one or another type of mutants capable to grow on it (polarized photon passed through). Again, the distribution of probabilities has nothing to do with adaptation, it is the set of outcomes that depends on the measurement conditions.

In spite of a seeming contradiction, one may say that the appearance of directed mutations is based on chance. It is a result of a random choice (as in the usual quantum-mechanical case where the photon either passes through or does not pass through polarizer). The set of outcomes, however, *is determined by external conditions* (or angle of the polarizer in the case of photon). That is why this phenomenon has an adaptive or directed character.

The notion of mixed state can also be used in the suggested approach - for taking into account spontaneous mutations, which can accumulate, for example, before plating. The situation should be described with a density matrix to take care of the additional uncertainty in our knowledge of cell's genetic sequence brought about by spontaneous mutations. Moreover, in the cases when the majority of the cells die after plating, it is the spontaneous mutations that predominate and lead to the non-Poissonian distribution in the fluctuation test (e.g., the appearance of "Ton" mutations, resistant to T1 phage, in the classical work by Luria & Delbruck (1943)). Thus, the proposed formalism can accommodate canonical Darwinian selection as well.

## 6. Discussion.

Nowadays all biology is baptized in Darwinism, which explains all evolutionary adaptation by natural selection. No doubt, the appearance of Darwinism was a great scientific event; it demonstrated that the enormous diversity of life forms, finely tuned to their surroundings, emerged as a result of a natural process. There is however, a more naive, "pagan" idea of adaptation which does not imply Darwinian selection at all. Importantly, it can be applied even to inanimate systems. A simplest example is provided by liquid *adapting* its shape to the shape of container. Purely physical principle (i.e., minimum of free energy) instead of Darwinian selection governs in this case.

However, as this paper implies, on a fundamental level, the concept of selection can underlie the "pagan" adaptation as well: quantum-mechanical states are selected that are eigenfunctions of either the energy operator or of an observable operator, depending what situation one is considering. Again, only the "fitted" variants survive in both of these cases, all others being destroyed by interference. Contrary to the Darwinian case, however, the selected variants do not preexist independently from each other prior to selection, they all are components of one state. In other words, the selection happens not in a usual population of objects, but in the "*population of virtual states*" of each single object. Thus, the indefinite variability postulated by Darwin is realized not in the offspring of an individual but in the individual itself. Hence, in spite of selection involved, this kind of adaptation, associated in this paper with wave function reduction, can be considered Lamarkian as well.

The problem of quantum measurement and wave function reduction is the most challenging in the foundations of quantum mechanics. This article is not intended to contribute to its solution. Rather, it is an attempt to point to an analogy between quantum measurement and the phenomena of adaptation, and apply this analogy to the particular case of directed mutations. The very possibility to look at two seemingly different phenomena (measurement and directed mutations) from one unifying point of view has a value in and of itself.

*6.1. Can the proposed hypothesis serve as an explanation?*

The hypothesis of superposition proposed in this article has a formal character and is deliberately neglectful of the examining of intracellular processes. Can it then be considered as an explanation? Here follow several arguments in favor of this possibility.

1. The hypothesis makes verifiable predictions. This question will be discussed later (6-3). The hypothesis is also *falsifiable*: it will have to be rejected if real 'molecular mechanisms' for directed mutations will be discovered.

2. As far as "the formal character" is concerned, it should be noted that one does not have to turn to the intricate details of a physical system for explanation of all of its principal characteristics. E.g., description of a gyroscope's motion does not require knowledge of its molecular structure: it can be made from iron, stone, wood etc. It is enough to know that the gyroscope is solid and then use conservation principles based on space and time symmetries.

The bacterial cell is a complex mesoscopic object. I believe, however, that just as in the previous example some aspects of its behavior can be explained without knowledge of particular molecular details. The phenomenon of directed mutations is rather universal: they may be both point mutations and deletions (and, perhaps, transpositions etc. (Stahl 1990)). Therefore I believe that its explanation should eventually be based on some general properties of the cell and its capability to propagate.

*6.2. What is wrong with the "Central Dogma of Molecular Biology"?*

Obviously, I consider the proposed description to be sufficient only for general explanation of the phenomenon of directed mutations. The analysis of each unique situation will be complicated by the peculiarities of the examined system (details of regulation, peculiarities of the starved state, etc.). It is necessary to reveal relationships between the formalism proposed and the current biochemical concepts of intracellular processes; otherwise it would be impossible to move from general principles to actual experimental systems.

What in the "Central Dogma of Molecular Biology" could be incorrect? I am convinced that the search for the interpretation of the formalism proposed in the paper will not lead to the discovery of a *new special mechanisms, in the usual meaning of molecular biology*. Rather, the transition to the examination of an elementary living object should lead to important changes in our notions concerning the very physics of the intracellular dynamics. Turning to the examination of a whole individual cell we must review our knowledge of how various events are correlated within it, and, in particular, exactly how the genome determines the cell's structure and functioning.

As more elaborate substantiation of this point will be published elsewhere, I will limit myself by a couple of further unsupported assertions.

The "Central Dogma of Molecular Biology" claims that the transfer of information "from gene to character" is unidirectional. It is necessary, however, to understand the physical meaning of this statement. The notion of unidirectionality implies *physical irreversibility*. This idea, in turn, corresponds with the usual molecular biological experience. This experience, however, is based either on the modeling of intracellular processes *in vitro*, or on the research of large quantities of cells *in vivo* (mainly in the exponential phase of growth). What occurs in a *single cell* is almost unknown. My opinion is that upon examination of a single cell we will see that the notion of the irreversibility of processes in the cell contradicts the conservation of cell's ordered state. In order to solve the problem we will have to reexamine the physical basis of our ideas about the intracellular dynamics. An account of the quantum theory principles will be necessary.

This revision will result in an idea that cell may be in a state when the processes in it are *physically reversible*. The idea of reversibility does not mean a possibility for some intracellular processes to go backwards (the "biochemical" meaning of reversibility). The only important consequence of reversibility (taken from a *physical* point of view) - is that there is no loss of information in reversible processes, so the system must remember what initial event has led to its present state. That is why the cell must "*remember*" what event on a genomic level has led to the appearance of incorrect and useful protein. This point reveals the initial incorrectness of the "Central Dogma of Molecular Biology".

The relationship of the idea of reversibility and that of superposition, proposed in the paper, is the following. Reversible (Schroedingerian) evolution of physical system maintains the state of superposition. It was suggested above that the process of genome expression can be described by reversible equations of quantum theory. Therefore, initial uncertainty on genetic level (say, base tautomery, or transition of proton from one place of a nucleotide to another), which can be described using the concept of superposition, will expand to the state of the whole cell; i.e. the whole cell should be described using the concept of superposition.

### *6. 3. The Schroedinger cat paradox and macroscopic superposition.*

This section will clarify how one should understand the idea of the cell in "state of superposition" and what relation it has to the famous Shroedinger cat paradox (Shroedinger 1928)

This relation exists only in the following respect. This paradox demonstrates that the quantum-mechanical description of nature is considered to be of the fundamental importance, and that the idea of the superposition of macroscopical states, however embarrassing it is, is taken seriously in the physical community. Numerous papers were devoted to the issue how to avoid this contradicting common sense deduction from quantum mechanics (see, for example, Omnes 1992).

However, as to the hypothesis itself, it does not contradict common sense - the states forming the superposition are proposed to be in principle *indiscernible*, and only the change in environment (which is equivalent to performing the measurement) can allow us to distinguish between them. This kind of superposition is absolutely legitimate from the common sense point of view. Consider, for example, a phonon in a solid: it is treated as an elementary particle, and the superposition concept can be applied to the treatment. However, in reality phonons do not exist: it is just a convenient method to *describe the*

*dynamics of the whole solid*. (i.e. how motions of the atoms in it are correlated with each other). Thus, the concept of superposition is applicable to the dynamic of a macroscopic object. It merely follows from the fact that the state vectors of macroscopic systems "live" in a linear Hilbert space and therefore can be presented as a superposition dependent on what basis we choose for the description of the system (see also section 4 on "preferred basis"). Undoubtedly, since the macroscopic states which one usually needs to describe are exactly those that form the preferred basis, the use of superposition concept remains as just a formal possibility for the majority of classical objects. However, this is not the case when one considers biological system able to adapt to its environment: upon the change in the environment, the preferred basis can also change. If this indeed happens, the old states will have to be presented as a superposition of the states of the new basis. Then, the "adaptation" of the system to the new environment will occur as a reduction of the old state vector to one of the new ones. This is the basis of the use of the superposition concept in this paper.

Considering a macroscopic object in the three dimensional physical space, how one is to understand the concept of superposition? It is important that superposition of macroscopic states does not have to manifest itself in the spatial distinction between the states; it could be just a correlation between parts of a physical system, separated on macroscopic distances. The possibility of such correlations is a consequence of the *nonseparability* principle. This principle itself follows from the fact that the Hilbert space of composite system is a tensor product of Hilbert spaces of its parts; therefore, the state of a composite system in general case can not be reduced to a product of states of its parts and should be presented as their linear combination:

$$|\Psi> = \Sigma\, a^i \Pi |\psi^i_j> \qquad (3)$$

The simplest example is provided by a pair of correlated electrons, spin of which is measured in the EPR type experiment. The state of such composite system is presented as a superposition of two components, which correspond to two possible spin values of first electron, say $|A\uparrow>$ and $|A\downarrow>$ and second electron, say, $|B\uparrow>$ and $|B\downarrow>$ in the following

combination:

$$|\Psi> = a|A\uparrow>|B\downarrow> + b|A\downarrow>|B\uparrow> \qquad (3)$$

, where, to satisfy the law of momentum conservation, the spins of the two electrons have opposite values for each component of the superposition.

Upon measurement of the spin of one electron (say, $|A\uparrow>$), the reduction of the wave function of the whole system occurs. This reduction is a nonlocal act, since it redefines the spin value of the second electron (in this case, $|B\downarrow>$, according to the (3)); and if the spins of both electrons are measured, this nonlocality reveals itself in correlations between the results of these measurements. Since the electrons can be located on macroscopic distances from each other, these correlations can be considered a manifestation of superposition of macroscopic states of the two-electron system.

The above arguments are relevant for any many-part system, especially for the systems with tight interactions between parts, such as condensed states of matter. As this paper implies, nonseparability (and possibility to use the concept of superposition) is essential for the phenomenon of life as well. This paper was not intended to suggest any specific molecular details of how the directed mutations occur. One may hypothesise, however, that in this case nonseparability manifests itself in correlations between events on DNA level (e.g. recognition of nucleotides by transcription and replication machinery) and on the level of the phenotype (e.g. enzyme activity), due to the physical reversibiliy of intracellular processes (section 6.2). These correlations make it impossible to describe mutation as just a change on the DNA level; it should be considered as an event occurring on the level of the whole cell. This event happens with certain probability when the cell is placed in new conditions which allow some of the components of the superposition irreversibly amplify and generate a colony.

*6.4. Predictions.*

The proposed approach considers the ability to mutate in directed manner as a natural property of any single cell. When it was first formulated (1991), the hypothesis predicted that the phenomenon can be reproduced in eukaryotic cells. This has been recently reported for yeast (Hall 1992b; Steele, Jinks-Robertson 1992). Yeast and bacteria, however, are single cellular organisms, which could develop this as a special mechanism for their adaptive needs. The hypothesis predicts that the phenomenon should be found in somatic cells of multi-cellular organisms. There is hardly any use for it there; to the contrary, it may represent danger for the whole organism.

The last prediction is relevant to the health related problem of oncogenesis. The last decade development in this field made it evident that so-called oncogenes are often cellular genes related to the control of cellular proliferation. Several genetic events are usually required for cell transformation. For example, inactivation of so called tumor suppressor genes are often required for tumorogenesis. Accumulation of these changes is supposed to take place owing to selection of "fitted" (and harmful for the organism) mutants, which allow the cell to proliferate (Novel 1976). The mutations, in turn, are presumed to happen accidentally or to be induced by various agents although not in a direct manner. I believe that the phenomenon of tumor progression directly concerns the issue discussed, because mutations leading to cell's proliferation take the main role in this case as well. The mutations may not only be selected from *preexisting* spontaneously generated pool, but their very emergence may be induced by appropriate conditions, allowing the mutated cell to proliferate. I believe it may be tested experimentally.

The general scheme of an experiment may be like the following. One must find an experimental system in which a single mutational event (point mutation, deletion, transposition, etc.) leads to the cell's ability to propagate in certain conditions. Activation of an oncogene, inactivation of an "anti-oncogene," or reversion to prototrophy can all be examples of such an event. Besides, there must be more stringent conditions, in which the same event does not lead to the cell's propagation (for example, because of absence of additional growth factors). My prediction is that the frequency of the mutational event must be higher in the first case than in the second one.

It must be noted that the above predictions would not *prove* the hypothesis proposed. Neither should they be considered as crucial experiments (in the sense of

Popper (1976)) *disproving* "the old paradigm". More "trivial" explanation can account for these predictions such as that human somatic cells have special mechanisms which allow them to change their genome directly for optimal propagation. There are, however, two arguments against this explanation. 1. The existence of special mechanisms favoring tumor cells' reproduction seems hardly possible: somatic cells do not exist separately in nature, and do not need to proliferate as fast as possible. 2. It is important that the proposed predictions do not *follow* from "the old paradigm" (though do not contradict it and still can be explained with the help of the hypothesis *ad hoc*). From this point of view (originally proposed by Lacatos (1971)), the suggested hypothesis also seems to be more preferable.

I would like to thank M.V.Volkenstein and E.A.Liberman for helpful discussion, K.M.Chumakov and G.V.Resapkin for discussion and encouragment and E.V.Koonin for discussion and help in the drafting of first version of the manuscript.

**Figure legends**

**Fig. 1. Two basic experiments on directed mutations**. (For more detailed description and controls see reviews by Foster (1992) and Hall (1994))

**A.** Fluctuation test of Luria and Delbruck.
Two models of mutant distribution in independent cultures are considered.
 **1.** Spontaneous generation of mutations any time before plating leads to large fluctuations (top part, left scenario and right scenario) from one culture to another and non-Poisson distribution (bottom part, lower dashed curve).
**2.** If the mutations emerge only after plating, less variation is generated (top part, right scenario only), leading to a Poisson distribution (bottom part, higher dashed curve). The Ryan and Cairns observed a distribution that was a composite of these two models (bottom part, solid curve) suggesting that part of mutations arise only after the cells were plated in the selective conditions.

**B.** Delayed application of selective medium.
This experiment is designed to rule out the effect of stationary state (starvation) on the mutation frequency.
Top part. The cells are plated on the plate without growth medium. At different times after plating the selective medium is added allowing the mutants to grow. The accumulation of mutants is observed for several days.
Bottom part. Two models are considered. 1. Mutation frequency is increased as a result of starvation. This predicts that the mutations will start to accumulate right after plating and upon application of selective medium all the mutants will give rise to colonies. (dashed curves). 2. Starvation does not increase the mutation frequency. In this case the whole time course of appearance of colonies is delayed one to three days, depending on the day of application of selective medium (solid curves).
The experimental results agreed with the prediction of the second model.

**Fig. 2. Comparison of the photon polarization experiment and the bacteria plating experiment.**

A. Two photons of the same polarization move through two differently oriented polarizers. In each case, the state vector $|\Psi>$ of each photon is presented as a superposition of two orthogonal components, and only one of them passes (survives) the polarizer. Reduction of the state vector occurs, in each case by selection of the "fit" variant.

B. Two cells of the same genotype and phenotype are plated on two plates with different selective mediums. In the top case the Lac+ mutation will give raise to a colony, in the bottom case the Trp+ mutation will give raise to a colony. The hypothesis states that in each case the state vector of the cell should be presented as a superposition of several components (on the figure two components, for the simplicity), determined by the plating conditions.

**Figure 1**

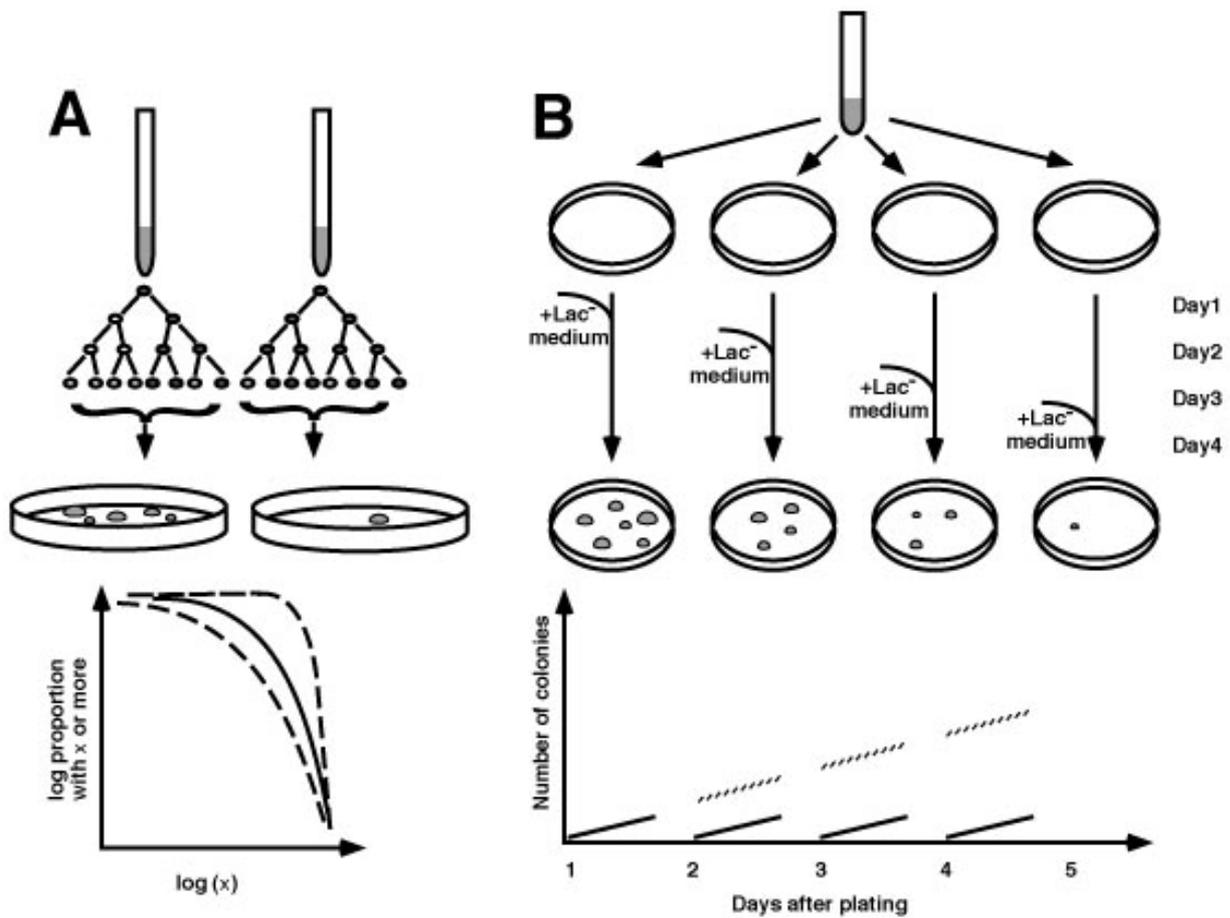

**Figure 2**

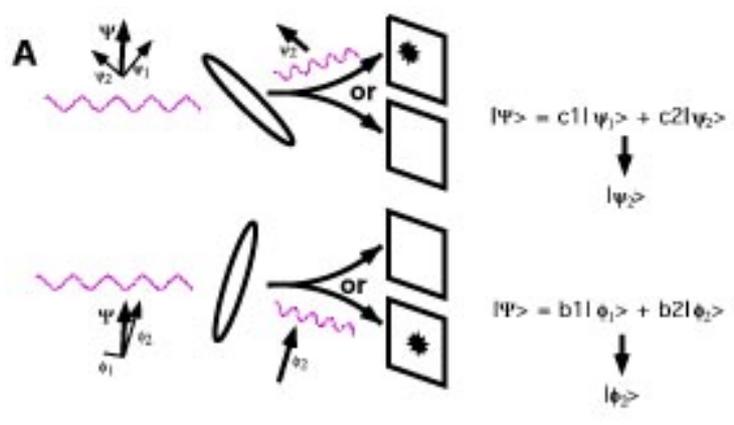
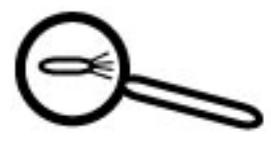
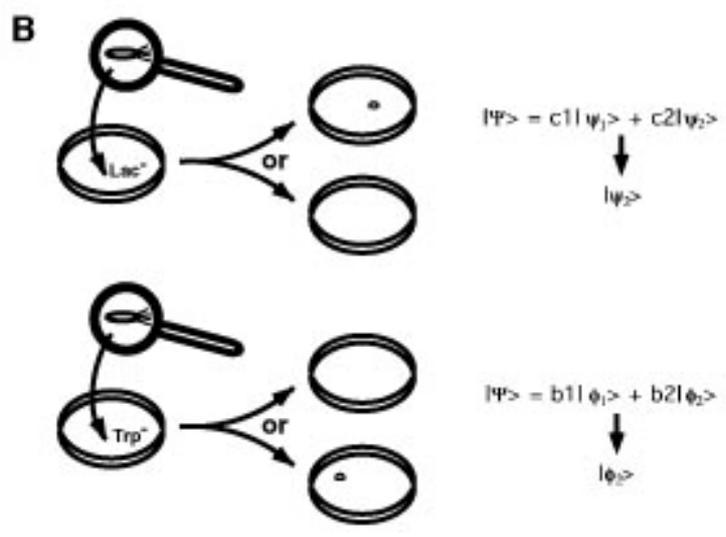